\input{aipcheck}

\documentclass[
    ,final            
  ]
  {aipproc}

\layoutstyle{6x9}

\begin{document}

\title{``Ab initio'' models of galaxy formation: successes and open problems}


\author{Gabriella De Lucia\footnote{Present address: INAF - Astronomical
  Observatory of Trieste, via Tiepolo 11, I-34143 Trieste, Italy}}{
  address={Max--Planck--Institut f\"ur Astrophysik, 
        Karl--Schwarzschild--Str. 1, D-85748 Garching, Germany}
}

\begin{abstract}
In the past decades, different approaches have been developed in order to link
the physical properties of galaxies to the dark matter haloes in which they
reside. In this review, I give a brief overview of methods, aims, and limits of
these techniques, with particular emphasis on semi-analytic models of galaxy
formation. For these models, I also provide a brief summary of recent successes
and open problems. 

\end{abstract}

\maketitle

\section{Introduction}

During the last decades, a number of observational tests have converged to
establish the $\Lambda$CDM model as the {\it de facto} standard cosmological
model for structure formation. Over the past years, it has been shown that this
model can account simultaneously for the present acceleration of cosmic
expansion as inferred from supernovae explosions \cite{Perlmutter_etal_1999},
the structure seen in the $z=3$ Ly$\alpha$ forest \cite{Mandelbaum_etal_2003},
the power spectrum of low redshift galaxies \cite{Percival_etal_2007}, the most
recent measurements of the microwave background fluctuations
\cite{Komatsu_etal_2008}, and a number of other important observational
constraints. Although the candidate particle for the non-baryonic dark matter
has yet to be detected in the laboratory, and the nature of dark energy remains
unknown, the fundamental cosmological parameters are now known with
uncertainties of only a few per cent, removing a large part of the parameter
space in galaxy formation studies.

While the basic theoretical paradigm for structure formation appears to be well
established, our understanding of the physical processes that lead to the
variety of observed galaxy properties is still far from complete. Although I
have kept the word {\it ab initio} in the title of this review, as suggested by
the organizers of this meeting, I would like to stress that {\it ab initio}
treatments of the galaxy formation process are very difficult - if not
unfeasible - simply because we do not have a complete understanding of the many
different and complex physical processes which are at play.

Today's models of galaxy formation find their seeds in the pioneering work by
\citet{White_Rees_1978} who proposed that galaxies form when gas condenses at
the centre of dark matter haloes, following the radiative cooling of
baryons. In the following years, three different approaches have been developed
in order to link the observed properties of luminous galaxies to the dark
matter haloes in which they reside. 

In {\bf semi-analytic models} of galaxy formation, which I will discuss in more
detail in the following, the evolution of the baryonic components of galaxies
is modelled using {\it simple} yet {\it physically and/or observationally
motivated prescriptions}. Modern semi-analytic models take advantage of
high-resolution N-body simulations to specify the location and evolution of
dark matter haloes, which are assumed to be the birth-places of luminous
galaxies. Since pure N-body simulations can handle very large number of
particles, this approach can access very large dynamic ranges in mass and
spatial resolution. In addition, the computational costs are limited so that
the method allows a fast exploration of the parameter space and an efficient
investigation of different specific physical assumption.

{\bf Direct hydrodynamical simulations} provide an explicit description of gas
dynamics. As a tool for studying galaxy formation, it is worth reminding that
these methods are still limited by relatively low mass and spatial resolution,
and by computational costs that are still prohibitive for simulations of
galaxies throughout large volumes. In addition, and perhaps more importantly,
complex physical processes such as star formation, feedback, etc. still need to
be modelled as {\it sub-grid physics}, either because the resolution of the
simulation becomes inadequate or because (and this is almost always true) we do
not have a `complete theory' for the particular physical process under
consideration.

A third approach - usually referred to as the {\bf Halo Occupation Distribution
(HOD) models} - has become popular in more recent years. This method
essentially bypasses any explicit modelling of the physical processes driving
galaxy formation and evolution, and specifies the link between dark matter
haloes and galaxies in a purely statistical fashion. The method is conceptually
very simple and easy to implement, and it can be constrained using the
increasing amount of available information on clustering properties of galaxies
at different cosmic epochs. It remains difficult, however, to move from a
purely statistical characterization of the link between dark matter haloes and
galaxies to a more physical understanding of the galaxy formation process
itself. 

Clearly, each of these methods has its own advantages and weaknesses, and they
should be viewed as {\it complementary rather than competitive}. In the
following, I will focus on semi-analytic models of galaxy formation. In
particular, I will provide a brief overview of the methods, aims, and limits of
these techniques, and give a brief summary of their recent successes and open
problems.

\section{Methods, aims, and limits}

The backbone of any semi-analytic model is provided by what in the jargon is
called a dark matter `merger tree', which essentially provides a representation
of the assembly history of a dark matter halo. Early renditions of
semi-analytic models - but this is still the case for a large number of
applications today - took advantage of the extended Press-Shechter (EPS)
formalism \cite{Bond_etal_1991, Bower_1991} and Monte Carlo methods to
construct representative histories of merger trees leading to the formation of
haloes of a given mass. It is important to note that some recent work has
demonstrated that this formalism might not provide an adequate description of
the merger trees extracted directly from numerical simulations
\cite{Benson_etal_2005,Li_etal_2007,Cole_etal_2008}. Although some of these
studies have provided `corrections' to analytic merger trees, many applications
are still carried out using the classical EPS formalism, and little work has
been done so far to understand to which measure this can affect predictions of
galaxy formation models.

As mentioned earlier, modern semi-analytic models (sometimes referred to as
{\it hybrid} models) take advantage of high-resolution N-body simulations to
follow the evolution of dark matter haloes in its full geometrical complexity
\cite{Kauffmann_etal_1999,Benson_etal_2000}. On a next level of complexity,
some more recent implementations have explicitly taken into account dark matter
substructures, i.e. the haloes within which galaxies form are still followed
when they are accreted onto a more massive system
\cite{Springel_etal_2001,DeLucia_etal_2004a}. There is one important caveat to
bear in mind regarding these methods: dark matter substructures are fragile
systems that are rapidly and efficiently destroyed below the resolution limit
of the simulation \cite{DeLucia_etal_2004b,Gao_etal_2004}. Since the baryons
are more concentrated than dark matter, it is to be expected that the baryonic
component will be more resistant to the tidal stripping that reduces its parent
halo mass. This creates a complex and strong position-dependent relation
between dark matter substructures and galaxies, contrary to what was assumed in
early HOD models. In addition, this treatment introduces a complication due to
the presence of `orphan galaxies', i.e. galaxies whose parent substructure mass
has been reduced below the resolution limit of the simulation. In most of the
available semi-analytic models, these galaxies are assumed to merge onto the
corresponding central galaxies after a residual merging time which is given by
some variation of the classical dynamical friction formula. Only a few models
account for the stripping of stars due to tidal interactions with the parent
halo.

Once the backbone of the model is constructed, using either N-body simulations
or analytic methods, galaxy formation and evolution is `coupled' to the merger
trees using a set of analytic laws that are based on theoretical and/or
observational arguments, to describe complex physical processes like star
formation, supernovae and AGN feedback processes, etc. Adopting this formalism,
it is possible to express the full galaxy formation process through a set of
differential equations that describe the variation in mass of the different
galactic components (e.g. gas, stars, metals). Given our limited understanding
of the physical processes at play, these equations contain {\it free
parameters} whose value is typically chosen in order to provide a reasonably
good agreement with observational data in the local Universe.

One common criticism to semi-analytic models is that there are {\it too many}
free parameters. It should be noted, however, that the number of these
parameters is not larger than the number of published comparisons with
different and independent sets of observational data, for any of the
semi-analytic models discussed in the recent literature. In addition, these are
not `statistical' parameters but, as explained above, they are due to our lack
of understanding of the physical processes considered. A change in any of these
parameters has consequences on a number of different predictable properties,
so that often there is little parameter degeneracy for a given set of
prescriptions. Finally, observations and theoretical arguments often provide
important constraints on the range of values that different parameters can
assume.

One great advantage of hybrid methods with respect to classical techniques
based on the EPS formalism, is that they provide full dynamical information
about model galaxies. Using this approach, it becomes possible to construct
realistic {\it mock catalogues} that contain not only physical information
about model galaxies such as masses, star formation rates, luminosities,
etc. but also dynamically consistent redshift and spatial information, like in
real redshift surveys. Using these mock catalogues, it is then possible to
carry-out detailed comparisons with observational data at different cosmic
epochs. These comparisons provide useful information on the relative importance
of different physical processes in establishing a certain observational trend,
and on the physics which is eventually missing in these models.

\section{Recent successes and open problems}

In discussing recent successes of semi-analytic models, I will start from the
most fundamental description of the galaxy population: the galaxy luminosity
function. Since early implementations of semi-analytic techniques, it was clear
that a relatively strong supernovae feedback was needed in order to suppress
the large excess of faint galaxies, due to the steep increase of low-mass dark
matter haloes \cite{White_and_Frenk_1991, Benson_etal_2003}. It is interesting
to note that matching the faint end of the luminosity function comes at the
expenses of exacerbating the excess of luminous bright objects, due to the fact
that the material reheated and/or ejected by low-mass galaxies ends up in the
hot gas associated to central galaxies of relatively massive haloes. At later
times, this material cools efficiently onto the corresponding central galaxies
increasing their luminosities and star formation rates, at odds with
observational data.

Matching the bright end of the luminosity function has proved difficult for a
long time, and a good match has been achieved only recently using a relatively
strong form of `radio-mode' AGN feedback
\cite{Croton_etal_2006,Bower_etal_2006}. Different prescriptions of AGN
feedback have been proposed, and still much work remains to be done in order to
understand if and how the energy injected by intermittent radio activity
at the cluster centre is able to efficiently suppress the cooling flows. Recent
observational measurements indicate that the ensemble-averaged power from radio
galaxies seems sufficient to offset the mean level of cooling
\cite{Best_etal_2007}. It is, however, important to note that not every cluster
shows central radio activity, and that the steep dependence of the radiative
cooling function on density makes it difficult to stabilize cooling flows at
all radii.

The main reason for the success of the `radio-mode' AGN feedback is that it is
not connected to star formation, so that its implementation permits at the same
time to suppress the luminosity of massive galaxies and to keep their stellar
populations old \cite{DeLucia_etal_2006}. Therefore these models seem to
reproduce, at least qualitatively, the observed trend for more massive
ellipticals to have shorter star formation time-scales. A good quantitative
agreement has not been shown yet and is complicated by large uncertainties
associated to star formation histories extracted from observed galaxy spectra. 

The suppression of late cooling (and therefore star formation) does not affect,
however, the assembly history of massive galaxies for which models predict an
increase in stellar mass by a factor 2 to 4 since $z\sim 1$, depending on
stellar mass \cite{DeLucia_etal_2006,DeLucia_and_Blaizot_2007}. This creates a
certain tension with the observation that the massive end of the galaxy mass
function does not appear to evolve significant over the same redshift interval.
Part of this tension is removed when taking into account observational errors
and uncertainties on galaxy mass estimates \cite[][see also Monaco these
proceedings]{Kitzbichler_and_White_2007,Stringer_etal_2008}. For the mass
assembly of the brightest cluster galaxies (BCGs), the situation is worse:
while observations seem consistent with no mass growth since $z\sim 1$, models
predict an increase in mass by a factor about 4
\cite{DeLucia_and_Blaizot_2007,Whiley_etal_2008}. One major caveat in this
comparison is given by the fact that observational studies usually adopt fixed
metric aperture magnitudes (which account for 25-50 per cent of the total light
contained in the BCG and intra-cluster light), while models compute total
magnitudes. Semi-analytic models do not provide information regarding the
spatial distribution of the BCG light, so aperture magnitudes cannot be
calculated. In addition, most of the available models do not take into account
the stripping of stars from other cluster galaxies due to tidal and harassment
effects \cite{Monaco_etal_2006, Conroy_etal_2007}.

Most of the models currently available exhibit a remarkable degree of agreement
with a large number of observations for the galaxy population in the local
Universe (e.g. the observed relations between stellar mass, gas mass, and
metallicity; the observed luminosity, colour, and morphology distribution; the
observed two-point correlation functions). When analysed in detail, however,
some of these comparisons show important and systematic (i.e. common to most of
the semi-analytic models discussed in the literature) disagreements.

Although models are not usually tuned to match observations of galaxy
clustering, they generally reproduce the observed dependence of clustering on
magnitude or colour. The agreement appears particularly good for the dependence
on luminosity, while the amplitude difference on colour appears greater in the
models than observed \cite{Springel_etal_2005,Coil_etal_2008}. This problem
might be (at least in part) related to the excess of small red satellite
galaxies which plagues all models discussed in the recent literature (e.g. see
Fig.~11 in \cite{Croton_etal_2006} and Fig.~4 in \cite{Somerville_etal_2008};
see also Monaco these proceedings).

A generic excess of intermediate to low-mass galaxies has been discussed by
\citet{Fontana_etal_2006}. At low redshift, this excess is largely due to
satellite galaxies that were formed and accreted early on, and that are
dominated by old stellar populations. Semi-analytic models assume that when a
galaxy is accreted onto a larger structure, the gas supply can no longer be
replenished by cooling that is suppressed by an instantaneous and complete
stripping of the hot gas reservoir. Since this process (commonly referred to as
`strangulation') is usually combined with a relatively efficient supernovae
feedback, galaxies that are accreted onto a larger system consume their gas
very rapidly, moving onto the red-sequence on quite short time-scales
\cite{Weinmann_etal_2006,DeLucia_2007,Wang_etal_2007}. This contributes to
produce an excess of faint and red satellites and a transition region
(sometimes referred to as `green valley') which does not appear to be as well
populated as observed.

Much effort has been recently devoted to this
problem. \citet{McCarthy_etal_2008} have used high resolution hydrodynamic
simulations to show that galaxies are able to retain a significant fraction of
their hot haloes following virial crossing. \citet{Font_etal_2008} incorporated
a simple model based on these simulations within the Durham semi-analytic
model. With this modification, a larger fraction of satellites has bluer
colours, resulting in a colour distribution that is in better (but not perfect)
agreement with observational data.

The completion of new high-redshift surveys has recently pushed comparisons
between model results and observational data to higher redshift. I do not have
time here to discuss in detail all agreements and disagreements discussed in
the recent literature. I would like to stress, however, that this still rather
unexplored regime for modern models is very interesting because it is at high
redshift that predictions from different models differ more dramatically.

To close this section, I would like to remind that a long standing problem for
hierarchical models has been to match the zero-point of the Tully-Fisher
relation (the observed correlation between the rotation speed and the
luminosity of spiral disks \cite{Tully_Fisher_1977}) while reproducing at the
same time the observed luminosity function. As discussed in \citet{Baugh_2006},
no model with a realistic calculation of galaxy size has been able to match the
zero-point of the Tully-Fisher relation using the circular velocity of the disk
measured at the half mass radius. It remains unclear if this difficulty is
related to some approximation in the size calculation, or if it is related to
more fundamental shortcomings of the cold dark matter model.

\section{Conclusions}

Given our poor knowledge of most of the physical processes at play, {\it ab
initio} treatments of the galaxy formation process are extremely difficult, if
not unfeasible. In the past decades, we have developed a number of {\it
techniques} to study galaxy formation within the currently standard
cosmological model. Semi-analytic models represent the most developed of these
techniques to make detailed predictions of galaxy properties at different
cosmic epochs.

These models are not meant to be definitive. Rather, they need to be {\it
falsified} against observational data, in order to gain insight on the relative
importance of different physical processes, and on the physics which is
eventually still missing in the models. When comparing model results with
observational data, it is important to take into account observational errors
and biases which are eventually introduced by a particular observational
selection and/or strategy. To this aim, realistic mock catalogues can be
constructed by coupling semi-analytic techniques with large cosmological N-body
simulations. Recently, a number of model results have been made publicly
available in the context of the modern concept of `Theoretical Virtual
Observatory'\footnote{A link to available galaxy catalogues and a documentation
can be found at http://www.mpa-garching.mpg.de/millennium/}. Considerable
interest has been shown by the astronomical community, and a large number of
papers using the public database have already been published, resulting in a
rapid refinement and verification of theoretical modelling.

The largest success of these techniques is to have shown that {\it we can study
galaxy formation within the currently established hierarchical paradigm}. The
largest failure is, unsurprisingly, that we have not yet solved the galaxy
formation problem. Undoubtedly, however, we have learnt a great deal about how
galaxies form and evolve, and how their physical properties are related to the
dark matter haloes in which they reside. 



\begin{theacknowledgments}
I wish to thank the organizers of the conference for the invitation, and for
the pleasant and stimulating atmosphere.
\end{theacknowledgments}



\bibliographystyle{aipproc}   

\bibliography{sample}

\IfFileExists{\jobname.bbl}{}
 {\typeout{}
  \typeout{******************************************}
  \typeout{** Please run "bibtex \jobname" to optain}
  \typeout{** the bibliography and then re-run LaTeX}
  \typeout{** twice to fix the references!}
  \typeout{******************************************}
  \typeout{}
 }

\end{document}